\pdfoutput=1
\newcommand*{\ATLASLATEXPATH}{}
\documentclass[PAPER, cernpreprint, texlive=2016, british]{\ATLASLATEXPATH atlasdoc}
 
\usepackage[subcaption]{\ATLASLATEXPATH atlaspackage}
\usepackage{\ATLASLATEXPATH atlasbiblatex}
 
\usepackage{empheq}
\usepackage{amsmath}
\usepackage{cleveref}
\usepackage{tabularx}
\usepackage{multirow}
 
\usepackage{\ATLASLATEXPATH atlasphysics}
 
\graphicspath{{logos/}{figures/}{figs/}}
 
\usepackage{ANA-EXOT-2018-05-PAPER-defs}

\AtlasTitle{Search for low-mass resonances decaying into two jets and produced in association with a photon using $pp$ collisions at $\sqrt{s} = 13$~\TeV\xspace with the ATLAS detector}
 
\AtlasAbstract{
A search is performed for localised excesses in dijet mass distributions of low-dijet-mass events produced in association with a high transverse energy photon.
The search uses up to 79.8\,\ifb\xspace of LHC proton--proton collisions collected by the ATLAS experiment at a centre-of-mass energy of $13$\,\TeV\xspace during 2015--2017.
Two variants are presented: one which makes no jet flavour requirements and one which requires both jets to be tagged as $b$-jets.
The observed mass distributions are consistent with multi-jet processes in the Standard Model.
The data are used to set upper limits on the production cross-section for a benchmark $Z^\prime$ model and, separately, on generic Gaussian-shape contributions to the mass distributions, extending the current ATLAS constraints on dijet resonances to the mass range between 225 and 1100 GeV.
}
 
\author{The ATLAS Collaboration}
 
\AtlasRefCode{EXOT-2018-05}
\PreprintIdNumber{CERN-EP-2018-347}
\AtlasJournal{Phys.\ Lett.\ B.}
\AtlasJournalRef{\PLB 795 (2019) 56}
\AtlasDOI{10.1016/j.physletb.2019.03.067}

\AtlasCoverSupportingNote{Search for resolved dijet resonances produced in association with a photon}{https://cds.cern.ch/record/2308306}
 
\AtlasCoverCommentsDeadline{24 December 2018}
 
\AtlasCoverAnalysisTeam{John Alison, Sarah Barnes, Antonio Boveia, Xin Chen, Eric Corrigan, Florencia Daneri, Roberta Devesa, Caterina Doglioni, Binbin Dong, Daniel Guest, Shih-Chieh Hsu, Adam Jinaru, William Kalderon, Karol Krizka, Zhiying Li, Samuel Meehan, Yvonne Ng, N. Nishu, Gustavo Otero Y Garzon, Katherine Pachal, Dianyu Liu, Emma Tolley, Alexandra Tudorache, Valentina Tudorache, Gang Zhang, Ning Zhou}

\AtlasCoverEdBoardMember{Claudia Beatriz Glasman Kuguel~(chair)}
\AtlasCoverEdBoardMember{Brigitte Vachon}
\AtlasCoverEdBoardMember{Miguel Villaplana Perez}
 
\AtlasCoverEgroupEditors{atlas-EXOT-2018-05-editors@cern.ch}
 
\AtlasCoverEgroupEdBoard{atlas-EXOT-2018-05-editorial-board@cern.ch}

\hypersetup{pdftitle={ATLAS document},pdfauthor={The ATLAS Collaboration}}
 
\begin{document}
 
\maketitle

\section{Introduction}
 
\let\oldTeV\TeV
\let\oldGeV\GeV
\renewcommand{\TeV}{\oldTeV\xspace}
\renewcommand{\GeV}{\oldGeV\xspace}
 
\newcommand{\PhotonGaussianLimitStatement}{from approximately 100--150~fb at a mean mass of 200~\GeV to approximately 10--20~fb at 1.2~\TeV}
\newcommand{\JetGaussianLimitStatement}{from approximately 100~fb at a mean mass of 350~\GeV to approximately 50~fb at 550~\GeV}
\newcommand{\JetZprimeLimitStatement}{\textbf{To be updated} for $\gq$ above 0.20 at mass $m_{\Zprime}=350$~\GeV and $\gq$ above 0.22 at $m_{\Zprime}=550$~\GeV}
 
Searches for resonant enhancements of the dijet invariant mass distribution (\mjj) are an essential part of the LHC physics programme.
New particles with sizeable couplings to quarks and gluons are predicted by many models, such as those including resonances with additional couplings to dark-matter particles~\cite{Chala:2015ama,LHCDMF:2015}.
 
Searches for dijet resonances with masses of several hundreds of \GeV to just above 1~\TeV have been carried out at lower-energy colliders~\cite{Arnison:1983dk,Albajar:1988rs,Bagnaia1984283, Aaltonen:2008dn,Alitti:1990kw} and at the LHC, which has also extended search sensitivities into the multi-$\TeV$ mass range~\cite{EXOT-2010-01,CMS-EXO-10-010,CMS-QCD-10-016,CMS-EXO-11-015,EXOT-2010-07,EXOT-2011-07,EXOT-2011-21,EXOT-2013-11,CMS-EXO-12-016,EXOT-2015-02,CMS-EXO-13-001,CMS-EXO-16-056,CMS-EXO-16-032,EXOT-2016-20,EXOT-2016-21}.
Despite using higher integrated luminosities than earlier colliders, these LHC searches have been limited at lower masses by a large multi-jet background.
Multi-jet events are produced at such high rates that fully recording every event would saturate the online data selection (called \textit{trigger}) and data acquisition systems.
To avoid this, minimum transverse momentum ($\ptmin$) thresholds are imposed on triggers collecting events with at least one jet (called single-jet triggers).
These thresholds create a lower bound on the sensitivity of searches
at a mass of approximately $\mjj \approx 2\ptmin$, where $\ptmin$ is typically several hundred $\GeV.$
Consequently, searches for dijet resonances at the LHC have poor sensitivity for masses below $1\,\TeV$, and set limits on the couplings of the resonance to quarks in this light-resonance region which are weaker than limits in heavy-resonance regions~\cite{Dobrescu:2013coa}.
Nevertheless, despite the difficulty of recording events containing light resonances, they remain a viable search target at the LHC, both from a model-agnostic point of view~\cite{Harris:2011bh} and, for example, in models of spin-dependent interactions of quarks with dark matter~\cite{Chala:2015ama,LHCDMF:2015}.
 
Recently, ATLAS and CMS have published searches for low-mass dijet resonances using several complementary strategies to avoid trigger limitations.
For $\mjj~>~450\,\GeV$, the most stringent limits are set by searches recording only partial event information~\cite{CMS-EXO-16-032,EXOT-2016-20}.
 
Another search avenue is opened by data in which a light resonance is boosted in the transverse direction via recoil against a high-$\pt$ photon~\cite{An:2012ue,Shimmin:2016vlc}.
Requiring a high-\pT photon in the final state reduces signal acceptance but allows efficient recording of events with lower dijet masses.
At even lower resonance masses, the decay products of the resonance will merge into a single large-radius jet.
Searches for this event signature have been used to set limits on resonant dijet production at both ATLAS~\cite{EXOT-2017-01} and CMS~\cite{CMS-EXO-17-001,Sirunyan:2018ikr}.
However, these searches become less sensitive above $200\,\GeV$--$350\,\GeV$, when the decay products fall outside the large-radius jet cone.
 
This Letter presents a new search for resonances in events containing a dijet and a high-\pT photon in the final state, using proton--proton ($pp$) collisions recorded at a centre-of-mass energy $\sqrt{s} =$ 13~\TeV and corresponding to an integrated luminosity up to 79.8~\ifb.
The search targets a dijet mass range of \ApproxMassRangeGamma.
This range covers masses below the range accessible using single-jet triggers or partial-event data and above the mass range where the resonance decay products merge.
The search is performed using samples of events selected either with or without criteria designed to identify jets originating from bottom quarks (\textit{$b$-jets}).
Searching in a subset of the data selected with $b$-jet identification criteria enhances sensitivity to resonances which preferentially decay into bottom quarks.
This search probes masses above $225\,\GeV$, obtaining results complementary to the reach of previous dijet searches at a centre-of-mass energy of $\sqrt{s} =$ 13~\TeV: below approximately $600\,\GeV$, previous ATLAS di-$b$-jet searches lose sensitivity~\cite{EXOT-2016-33}, while the range of the CMS boosted di-$b$-jet search~\cite{Sirunyan:2018ikr} is limited to a mass region up to 350~\GeV. Another complementary CMS search for resonances with masses above $325\,\GeV$ decaying to $b$-jets at a centre-of-mass energy of $\sqrt{s} =$ 8~\TeV is described in Ref.~\cite{CMS-EXO-16-057}.

\section{ATLAS detector}

\newcommand{\AtlasCoordFootnote}{
ATLAS uses a right-handed coordinate system with its origin at the nominal interaction point (IP) in the centre of the detector and the $z$-axis along the beam pipe.
The $x$-axis points from the IP to the centre of the LHC ring, and the $y$-axis points upwards.
Cylindrical coordinates $(r,\phi)$ are used in the transverse plane, with $\phi$ being the azimuthal angle around the $z$-axis.
The pseudorapidity is defined in terms of the polar angle $\theta$ as $\eta = -\ln \tan(\theta/2)$.
It is equivalent to the rapidity for massless particles.
Transverse momentum and energy are defined as $\pt\equiv p \sin{\theta}$ and $\ET\equiv E \sin{\theta}$, respectively.
Angular distance is measured in units of $\Delta R \equiv \sqrt{(\Delta\eta)^{2} + (\Delta\phi)^{2}}$.}
 
The ATLAS experiment~\cite{PERF-2007-01,Capeans:1291633,CERN-LHCC-2012-009,Abbott:2018ikt} at the LHC is a multipurpose particle detector with a forward--backward symmetric cylindrical geometry\footnote{\AtlasCoordFootnote\xspace} with layers of tracking, calorimeter, and muon detectors over nearly the entire solid angle around the $pp$ collision point.
The directions and energies of high transverse momentum particles are measured using tracking detectors, finely segmented hadronic and electromagnetic calorimeters, and a muon spectrometer, within axial and toroidal magnetic fields.
The inner tracker consists of silicon pixel, silicon microstrip, and transition radiation tracking detectors, and reconstructs charged-particle tracks in $|\eta| < 2.5$.
Lead/liquid-argon (LAr) sampling calorimeters provide electromagnetic (EM) energy measurements with high granularity.
A steel/scintillator-tile hadronic calorimeter covers the central pseudorapidity range ($|\eta| < 1.7$).
The endcap and forward regions are instrumented with LAr calorimeters for EM and hadronic energy measurements up to $|\eta| = 4.9$.
The trigger system~\cite{TRIG-2016-01} consists of a first-level trigger implemented in hardware, using a subset of the detector information to reduce the accepted rate to 100 kHz, followed by a software-based trigger that reduces the rate of recorded events to about 1 kHz.
 
\section{Data samples and event selection}

The result presented in this Letter is based on data collected in $pp$ collisions at $\sqrt{s} = 13\,\TeV$ during 2015--2017.
The signal consists of events with two jets from the decay of a new particle, and an additional photon, radiated off one of the colliding partons.
 
Data were collected via either a single-photon trigger or a combined trigger requiring additional jets, to allow a lower \pT requirement on the photon.
The data collected with the single-photon trigger are used to search for resonances with masses from 225 GeV to 450 GeV, while the data collected with the combined trigger are used to search for resonances with masses from 450 GeV to 1.1 TeV.
 
The single-photon trigger requires at least one photon candidate with $\photonPtTrig > 140\,\GeV$, where $\photonPtTrig$
is the photon transverse energy as reconstructed by the software-based trigger.
The combined trigger requires a photon and two additional jet candidates, each with $\pt > 50\,\GeV$.
The combined trigger requires $\photonPtTrig > 75\,\GeV$ for the 2016 data, increasing to $\photonPtTrig > 85\,\GeV$ for the 2017 data.
This trigger was not active during the 2015 data-taking period.
As a consequence, the single-photon trigger recorded $79.8\,\ifb$ of data and the combined trigger recorded $76.6\,\ifb$ of data.
Both triggers are fully efficient within uncertainties in the kinematic regimes used for this analysis.
 
After recording the data, a subset of collision events consistent with the signal are selected to populate $\mjj$ distributions for subsequent analysis.
A brief description of the reconstruction methods is given below together with the event selection.
 
In all of the events selected for analysis, all components of the detector are required to be operating correctly.
In addition, all events are required to have a reconstructed primary vertex~\cite{ATLAS-CONF-2014-018}, defined as a vertex with at least two reconstructed tracks, each with $\pT > 500~\MeV$.
 
Photon candidates are reconstructed from clusters of energy deposits in the electromagnetic calorimeter~\cite{PERF-2017-02}.
The energy of the candidate is corrected by applying energy scale factors measured with $Z \rightarrow e^+e^-$ decays~\cite{PERF-2013-05}.
 
The trajectory of the photon is reconstructed using the longitudinal segmentation of the calorimeters along the shower axis (shower depth) and a constraint from the average collision point of the proton beams.
Candidates are restricted to the region $|\eta| < 2.37$, excluding the transition region $1.37 < |\eta| < 1.52$ between the barrel and endcap calorimeters to ensure that they arise from well-calibrated regions of the calorimeter. An additional requirement is applied on the transverse energy of the photon candidate after reconstruction, which is required to have $\photonPt > 95\,\GeV$, where $\photonPt$ is the  transverse energy of the photon candidate after reconstruction.
 
Quality requirements are applied to the photon candidates to reject events containing misreconstructed photons arising from instrumental problems or from non-collision backgrounds.
Further \emph{tight} identification requirements are applied to reduce contamination from $\pi^0$ or other neutral hadrons decaying into two photons~\cite{PERF-2017-02}.
The photon identification is based on the profile of the energy deposits in the first and second layers of the electromagnetic calorimeter.
In addition to the tight identification requirement, candidates must meet \emph{tight isolation} criteria using calorimeter and tracking information, requiring that they be separated from nearby event activity~\cite{PERF-2017-03,HIGG-2016-17}.
Converted photon candidates matched to one track or a pair of tracks passing inner-detector quality requirements~\cite{PERF-2017-02} and satisfying tight identification and isolation criteria are also considered.
Any pair of matching tracks must form a vertex that is consistent with originating from a massless particle.
 
Jets are reconstructed using the anti-$k_{t}$ algorithm~\cite{antikT,Cacciari:2006} with radius parameter $R = 0.4$ from clusters of energy deposits in the calorimeters~\cite{PERF-2014-07}.
Quality requirements are applied to remove events containing spurious jets from detector noise and out-of-time energy deposits in the calorimeter from cosmic rays or other non-collision sources~\cite{ATLAS-CONF-2015-029}.
Jet energies are calibrated to the scale of the constituent particles of the jet and corrected for the presence of multiple simultaneous (pile-up) interactions~\cite{PERF-2014-03,PERF-2016-04}.
 
After reconstruction, jets with transverse momentum $\jetPt > 25\,\GeV$ and rapidity $|\jetEta| < 2.8$ are considered.
To suppress pile-up contributions, jets with $\jetPt < 60\,\GeV$ and $|\jetEta|<2.4$
are required to originate from the primary interaction vertex with the highest summed $\pT^2$ of associated tracks.
If a jet and a photon candidate are within $\Delta R = 0.4$, the jet candidate is removed.
 
These requirements retain approximately 30\% of a typical signal sample.
 
Jets which likely contain $b$-hadrons are identified (\btagged) with the \textsc{DL1} flavour tagger~\cite{ATL-PHYS-PUB-2017-013}.
Tracks are selected in a cone around the jet axis, using a radius which shrinks with increasing \jetPt.
The selected tracks are used as input to algorithms which attempt to reconstruct a $b$-hadron decay chain.
The resulting information is passed to a neural network which assigns a $b$-jet probability to each jet.
To account for mismodelling in simulated $b$-hadron decays, a comparison of the discrimination power of this network in data and Monte Carlo simulation is performed and correction factors are applied to simulation to reproduce the data~\cite{PERF-2016-05}.
Jets are considered \btagged when the \textsc{DL1} score exceeds a threshold consistent with a 77\% $b$-hadron identification efficiency on a benchmark $t \bar{t}$ sample. At this threshold, only 0.7\% light-flavour jets and 25\% charm-jets are retained.
 
Events which contain at least one photon candidate and two jets are selected using the above criteria
and separated into four categories for further analysis.
Two of the categories are constructed with flavour-inclusive criteria, for which \btagging\ results are ignored.
One of these two categories contains events recorded via the single-photon trigger, and the other category contains events recorded via the combined trigger.
To ensure the trigger is fully efficient, events in the single-photon-trigger category are required to have a photon with $\photonPt >150\,\GeV$ and events in the combined-trigger category are required to have a photon with $\photonPt > 95\,\GeV$ and two jets with $\jetPt > 65\,\GeV$.
The remaining two categories consist of events selected as in the flavour-inclusive categories, except that the two highest-\jetPt jets must satisfy the $b$-tagging criteria and have $|\jetEta| < 2.5$ to ensure that they fall within the acceptance of the tracking detectors.
 
Dijet production at the LHC occurs largely via $t$-channel processes, leading to jet pairs with high absolute values of $\yStar = (y_1-y_2)/2$, where $y_1$ and $y_2$ are the rapidities of the highest-\pT\ (leading) and second-highest-\pT\ (subleading) jet, respectively.
On the other hand, heavy particles tend to decay more isotropically, with the two jets having lower $|\yStar|$ values.
Therefore, $|y^*| < 0.75$ is required for all four categories.
This selection rejects up to 80\% of the multi-jet background events while accepting up to 80\% of the signal events discussed below.
A further selection is applied to select events above a given invariant mass depending on the trigger, $\mjj > 169~\GeV$ for the single-photon trigger and $\mjj > 335~\GeV$ for the combined trigger.
This is so that the background can be described by a smoothly falling analytic function satisfying the goodness-of-fit criteria described in~\ref{sec:background}.

\begin{table}[!h]
\setlength{\tabcolsep}{10pt}
\centering
\caption[]{Event selections used to construct each of the four event categories, as described in the text.}
\begin{tabular}{ l c c c c}
\toprule
Criterion & Single-photon trigger & Combined trigger \\
\midrule
Number of jets & \multicolumn{2}{c}{$\njet \geq 2$} \\
Number of photons & \multicolumn{2}{c}{$\nphoton \geq 1$} \\
Leading photon & $\photonPt > 150~\GeV$ & $\photonPt >  95~\GeV$ \\
Leading, subleading jet & $\jetPt > 25~\GeV$ & $\jetPt> 65~\GeV$ \\
Centrality & \multicolumn{2}{c}{$|\yStar|=|y_{1} - y_{2}|/2 < 0.75$}  \\
Invariant mass & $\mjj > 169~\GeV$ & $\mjj > 335~\GeV$ \\
\midrule
\midrule
Criterion (applied to each trigger selection) & Inclusive & $b$-tagged \\
\midrule
Jet $|\eta|$ & $|\jetEta| < 2.8$ & $|\jetEta| < 2.5$ \\
$b$-tagging & -- & $\ntag\geq 2$ \\
\bottomrule
\end{tabular}
\label{tab:analysisselection}
\end{table}
 
The above selections, summarised in Table~\ref{tab:analysisselection}, yield 2,522,549 and 15,557 events acquired by the single-photon trigger for the flavour-inclusive and \btagged categories, respectively.
They yield 1,520,114 and 9,015 events acquired by the combined trigger in the corresponding categories.
 
The distributions of $\mjj$ for events in each of the four categories are shown in Fig.~\ref{fig:data}.
Hypothetical signals with $m_{\Zprime}= 250\,\GeV$ and $m_{\Zprime}= 550\,\GeV$, as further discussed in Section~\ref{sec:limits}, are overlaid.
 
At the largest dijet masses considered, the combined-trigger categories provide greater sensitivity to signals than the single-photon-trigger categories due to their greater signal acceptance. The sensitivity is defined as $S/\sqrt{B}$, where $S$ and $B$ are the number of signal and background events in the simulation samples described in Section~\ref{sec:limits}.
At the smallest dijet masses considered, the jet $\pT$ thresholds of the combined trigger cause those categories to lose efficiency for signals and bias the $\mjj$ distributions of the background processes.
Therefore, to optimise the search across a wide range of signal masses, the invariant mass spectra selected using the combined-trigger categories are used in the search for signal masses above $450\,\GeV$, while the spectra obtained with the single-photon trigger are used for lower masses.
 
\section{Background estimation}
\label{sec:background}
 
To estimate the Standard Model contributions to the distributions in Fig.~\ref{fig:data}, smooth functions are fit to the data.
The dijet searches of the CDF, CMS, and ATLAS experiments~\cite{Aaltonen:2008dn,EXOT-2010-01,CMS-EXO-11-015,EXOT-2013-11,EXOT-2015-02,EXOT-2015-02,EXOT-2013-11,Alitti:1990kw,CMS-EXO-16-032} have successfully modelled dijet mass distributions in hadron colliders using a single function over the entire mass range considered in those searches.
This approach is not suitable when data constrain the fit too tightly for a single function to reliably model both ends of the distribution simultaneously.
Here, a more flexible technique is adopted, similar to that used in recent ATLAS dijet resonance searches~\cite{EXOT-2016-21,EXOT-2016-20}.
In this technique, a single fit using a given function over the entire mass distribution is replaced by many successive fits.
For each bin of the mass distribution, the same function is used to fit a broad mass range centred on the bin, and the background prediction for that bin is taken to be the value of the fitted function in the centre of the range.
The process is repeated for each bin of the mass distribution and the results are combined to form a background prediction covering the entire distribution. For invariant masses higher than the \mjj\ range used for the search (above 1.1 TeV), the window is allowed to extend beyond the range as long as data is available.
 
A set of parametric functions are considered for these fits:
 
\begin{equation}
f(x) = p_1 x^{-p_2} \mathrm{e}^{-  p_3  x - p_4  x^2}
\label{eq:ua2}
\end{equation}
 
or
 
\begin{equation}
f(x) = p_1 (1 - x)^{p_2} x^{p_3 + p_4\ln x  + p_5(\ln x)^2} \label{eq:5par},
\end{equation}
 
\noindent
where $x = \mjj/\sqrt{s}$ and $p_i$ are free parameters determined by fitting the $\mjj$ distribution.
In addition to the five-parameter function in Eq.~(\ref{eq:5par}), a four-parameter variant with $p_5 = 0$ and a three-parameter variant with $p_5 = p_4 = 0$ are also considered.
The width of the mass range used for the individual fits was optimised to retain the broadest possible range while maintaining a $\chi^2$ $p$-value above 0.05 in regions of the distribution that do not contain narrow excesses, where excesses are identified using the \BumpHunter algorithm described in the next section.
The sliding window procedure cannot be extended beyond the lower edge of the \mjj range used in each signal selection.
Therefore, until the optimal number of bins is reached on each side of a given bin center, the start of the window is fixed to the lower edge of the spectrum and the fitted functional form is evaluated for each bin in turn.
This procedure allows for a stable background estimate while maintaining sensitivity to signals localised in the \mjj distribution.
Tests performed by adding sample signals to smooth pseudo-data distributions confirmed that this approach can find signals of width-to-mass ratios up to 15\%, with sensitivity increasing for narrower signals.  The ranges of the individual fits vary from 750~\GeV in the narrowest case to 1600~\GeV in the widest case. A signal with a 15\% width-to-mass ratio constrained by the narrowest fit would have an absolute width of 163~\GeV, or less than one quarter of the fit range.
 
Monte Carlo samples of background containing a photon with associated jets were simulated using \textsc{Sherpa}~2.1.1 \cite{Gleisberg:2008ta}, generated in several bins of photon transverse momentum at the particle level (termed as $\photonPt$ for this paragraph), from 35~\GeV up to energies where backgrounds become negligible in data, at approximately 4~\TeV.
The matrix elements, calculated at next-to-leading order (NLO) with up to three partons for  $\photonPt<70$~\GeV or four partons for higher $\photonPt$,
were merged with the \textsc{Sherpa} parton shower~\cite{Schumann:2007mg} using the \textsc{ME+PS@LO} prescription~\cite{Hoeche:2009rj}.
The \textsc{CT10} set of parton distribution functions (PDF)~\cite{Lai:2010vv} was used in conjunction with the dedicated parton shower tuning developed by the \textsc{Sherpa} authors.
These samples, alone and in combination with the signal samples discussed below, were used to validate the background model obtained with the above mentioned method, and they were also used to verify that the fitting procedure is robust against false positive signals. Additionally, the simulated samples were used to calculate the fractional dijet mass resolution, which was found to be in the range 8\%--3\% for the masses of 225~\GeV up to 1.1~\TeV considered in this search.

\begin{figure}
\centering
\begin{subfigure}[b]{0.49\textwidth}
\includegraphics[width=\textwidth]{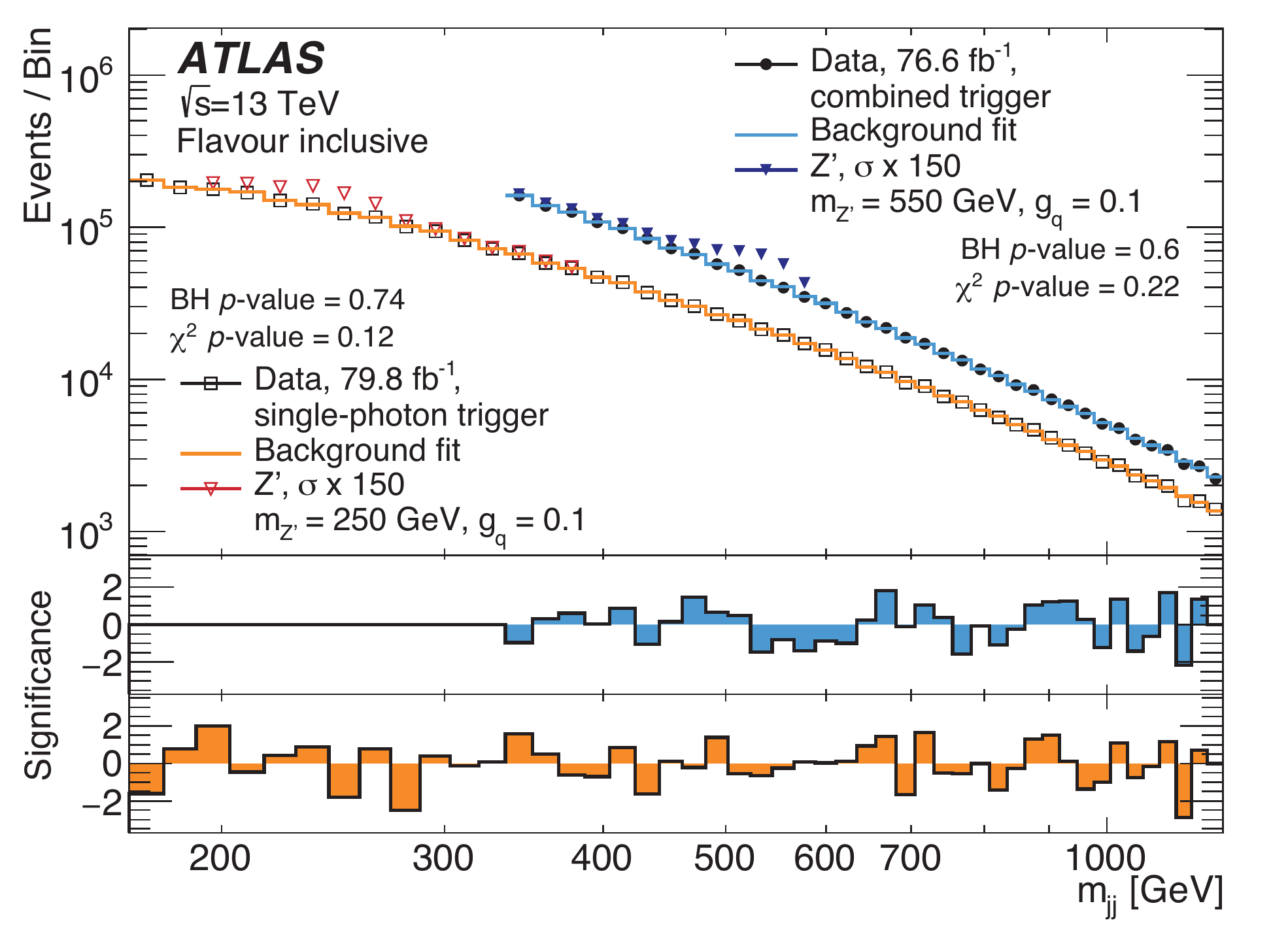}
\caption{\label{fig:1a}}
\end{subfigure}
\begin{subfigure}[b]{0.49\textwidth}
\includegraphics[width=\textwidth]{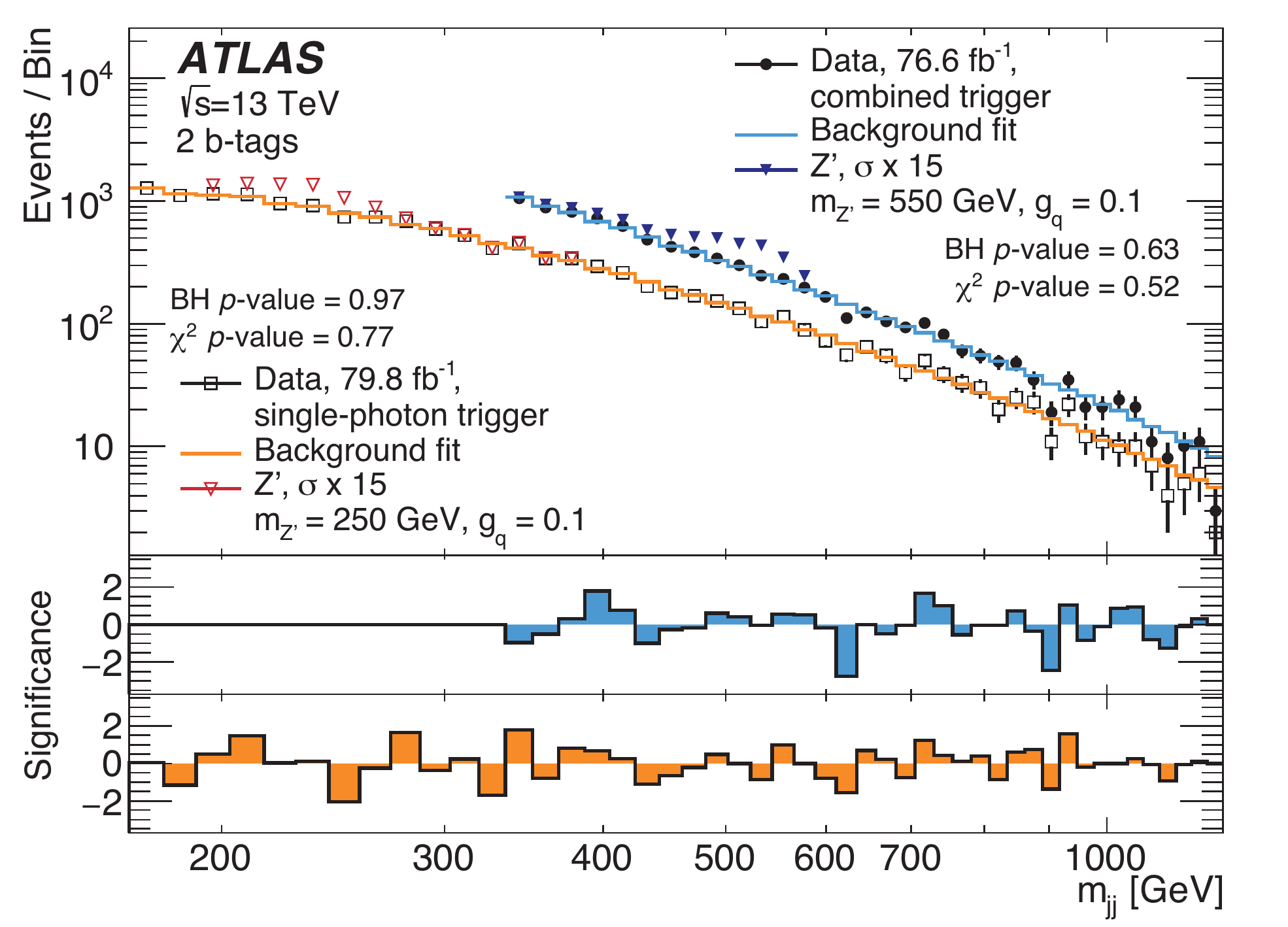}
\caption{\label{fig:1b}}
\end{subfigure}
\caption[]{Dijet mass distributions for the \subref{fig:1a} flavour-inclusive and \subref{fig:1b} \btagged categories.
In both figures, the distribution for the sample collected using the combined trigger with $\photonPt > 95\,\GeV$ and two $\jetPt > 25\,\GeV$ jets (filled circles) and the distribution for the sample collected using the single-photon trigger with $\photonPt > 150\,\GeV$  (open squares) are shown separately.
The solid lines indicate the background estimated from the fitting method described in the text.
Also shown are
the $p$-values  both by a $\chi^2$ comparison of data to background estimate and by \BumpHunter (BH).
The solid and empty triangles represent a $\Zprime$ injected signal with \gq = 0.1, masses of 550 and 250~\GeV, respectively, where the theory-cross section is multiplied by the factor shown in the legend.
The bottom panels show the significances of bin-by-bin differences between the data and the fits for the combined trigger (middle) and single-photon trigger (bottom).
These Gaussian significances are calculated from the Poisson probability, considering only statistical uncertainties on the data.
}
\label{fig:data}
\end{figure}

\section{Search results}
\label{sec:result}
 
Fig.~\ref{fig:data} shows the results of fitting each of the observed distributions, as described in Section~\ref{sec:background}.
For each distribution, the function among those in Eqs.~(\ref{eq:ua2}) and (\ref{eq:5par}) and their variants which yields the highest $\chi^2$ $p$-value (shown in the figure), in absence of localized excesses, is chosen as the primary function for the fitting method.
The function with the lowest $\chi^2$ $p$-value which still results in a $p$-value larger than 0.05 is chosen as an alternative function.
The primary and alternative functions for each of the four search categories are shown in Table~\ref{tab:fitsummary}.
The alternative function is used to estimate the systematic uncertainty of the background prediction due to the choice of function, as described below.

\begin{table}
\caption[]{Summary of functions used for background fits to each category.
The five-parameter function (5~par.) is given in Eq.~(\ref{eq:5par}).
The four-parameter variant (4~par.) sets $p_5 = 0$, while the three-parameter variant (3~par) sets $p_5 = p_4 = 0$.}
\begin{tabularx}{\textwidth}{ l | *4{>{\centering\arraybackslash}X}@{}}
\toprule
\centering
Fit               						        & Flavour-inclusive, single~$\gamma$~trigger     & Flavour-inclusive, combined~trigger 			& \btagged, single~$\gamma$~trigger   		   & \btagged, combined~trigger.        \\ \midrule
Primary fit 								    & Eq.~(\ref{eq:5par}), 5~par.   & Eq.~(\ref{eq:5par}), 4~par. & Eq.~(\ref{eq:5par}), 4~par. & Eq.~(\ref{eq:5par}), 3~par. \\
($\chi^2$ $p$-value)       						& (0.11) 					    & (0.23)  					  &  (0.75)  				    & (0.53) 					  \\ \midrule
Alternative fit 							    & Eq.~(\ref{eq:5par}), 4~par.	& Eq.~(\ref{eq:ua2}) 		  & Eq.~(\ref{eq:5par}), 3~par. & Eq.~(\ref{eq:5par}), 5~par.\\
($\chi^2$ $p$-value)       						& (0.07) 					    & (0.20)  					  &  (0.75)  				    & (0.44) 					  \\ \bottomrule
\end{tabularx}
\label{tab:fitsummary}
\end{table}
 
The statistical significance of any localised excess in each $\mjj$ distribution is quantified using the \BumpHunter~(BH) algorithm~\cite{Aaltonen:2008vt,Choudalakis:2011bh}.
The algorithm compares the binned $\mjj$ distribution of the data with the fitted background estimate, considering mass intervals centered in each bin location and with widths of variable size from two bins up to half the mass range used for the search (169 or 335 GeV to 1.1 TeV, for the single and combined trigger respectively).
 
The statistical significance of the outcome is evaluated using the ensemble of possible outcomes by applying the algorithm to many pseudo-data samples drawn randomly from the background fit.
Without including systematic uncertainties, the \BumpHunter\ $p$-value -- the probability that fluctuations of the background model would produce an excess at least as significant as the one observed in the data, anywhere in the distribution -- is $p > 0.5$ for all distributions.
Thus, there is no evidence of a localised contribution to the mass distribution from new phenomena.
 
\section{Limit setting}
\label{sec:limits}
 
Limits are set on the possible contributions to the $\mjj$ distributions from two kinds of resonant signal processes.
As a specific benchmark signal, a leptophobic $\Zprime$ resonance is simulated as in Refs.~\cite{LHCDMF:2015,EXOT-2015-02}.
The $\Zprime$ resonance has axial-vector couplings to quarks and to a fermion dark-matter candidate.
The coupling of the $\Zprime$ to quarks, $\gq$, is set to be universal in quark flavour.
The mass of the dark-matter fermion is set to a value much heavier than the $\Zprime$, such that the decay width to dark matter is zero.
The total width $\Gamma_{\Zprime}$ is computed as the minimum width allowed given the coupling and mass $m_{\Zprime}$; this width is $3.6\%$--$4.2\%$ of the mass for $m_{\Zprime}=0.25$--$0.95$~\TeV and $\gq=0.3$.
The interference between the $\Zprime$ in this benchmark model and the Standard Model $Z$ boson is assumed to be negligible.
A set of event samples were generated at leading order with $m_{\Zprime}$ values in the range 0.25--1.5~\TeV and with $\gq=0.3$ using \MGMCatNLOV{2.2.3}~\cite{Alwall:2014hca}; the \textsc{NNPDF3.0 LO} PDF set~\cite{Ball:2012cx} was used in conjunction with \PYTHIA~8.186~\cite{Sjostrand:2007gs} and the \textsc{A14} set of tuned parameters~\cite{ATL-PHYS-PUB-2014-021}.
For these samples, the acceptances of the kinematic selections in the flavour-inclusive categories range from 1\% to 2.5\%, increasing with signal mass, for the sample collected by the combined trigger and from 4\% to 10\% for the sample collected by the single-photon trigger.
For the \btagged categories, the kinematic acceptance is defined relative to the full flavour-inclusive generated samples, leading to acceptance values of 0.2\%--0.4\% and 0.7\%--1.6\% for the combined and single-photon trigger, respectively.
The reconstruction efficiencies range from 74\% to 80\% for the flavour-inclusive categories and from 40\% to 48\% for the \btagged categories, decreasing with increasing signal mass.
 
Limits are set on the considered new-physics contributions to the $\mjj$ distributions using a Bayesian method.
A constant prior is used for the signal cross-section and Gaussian priors for nuisance parameters corresponding to systematic uncertainties.
The expected limits are calculated using pseudo-experiments generated from the background-only component of a signal-plus-background fit to the data, using the same fitting ranges and functions selected as the best model in the search phase.
Signal hypotheses at discrete mass values are used to set 95\% credibility-level (CL) upper limits on the cross-section times acceptance ~\cite{EXOT-2010-07}.
The limits are obtained for a discrete set of points in the $\gq$--$m_{\Zprime}$ plane, shown in Fig.~\ref{fig:limits_zprime}.
 
\begin{figure}
\centering
\begin{subfigure}[b]{0.49\textwidth}
\includegraphics[width=\textwidth]{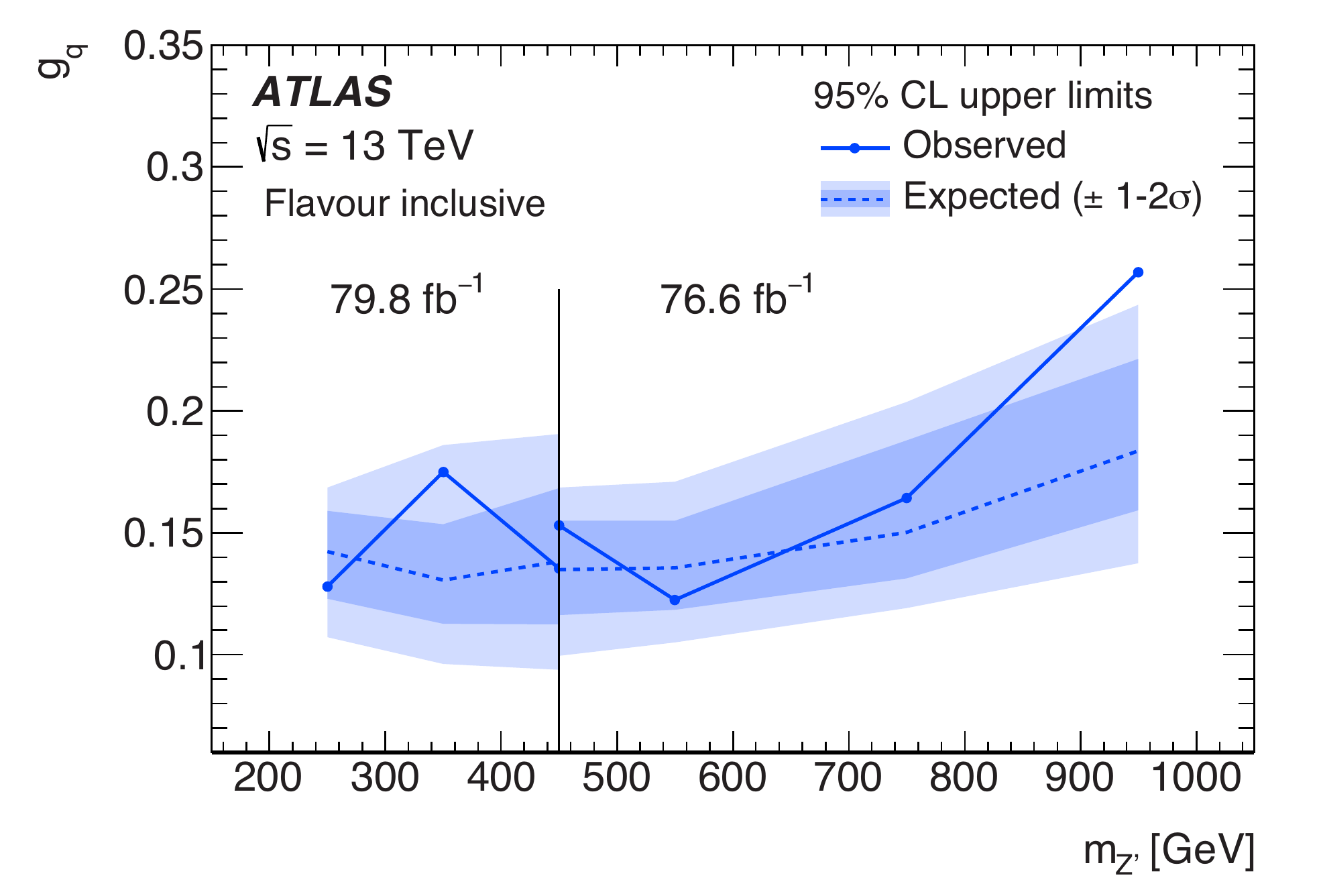}
\caption{\label{fig:limit_single_inc_0p2}}
\end{subfigure}
\begin{subfigure}[b]{0.49\textwidth}
\includegraphics[width=\textwidth]{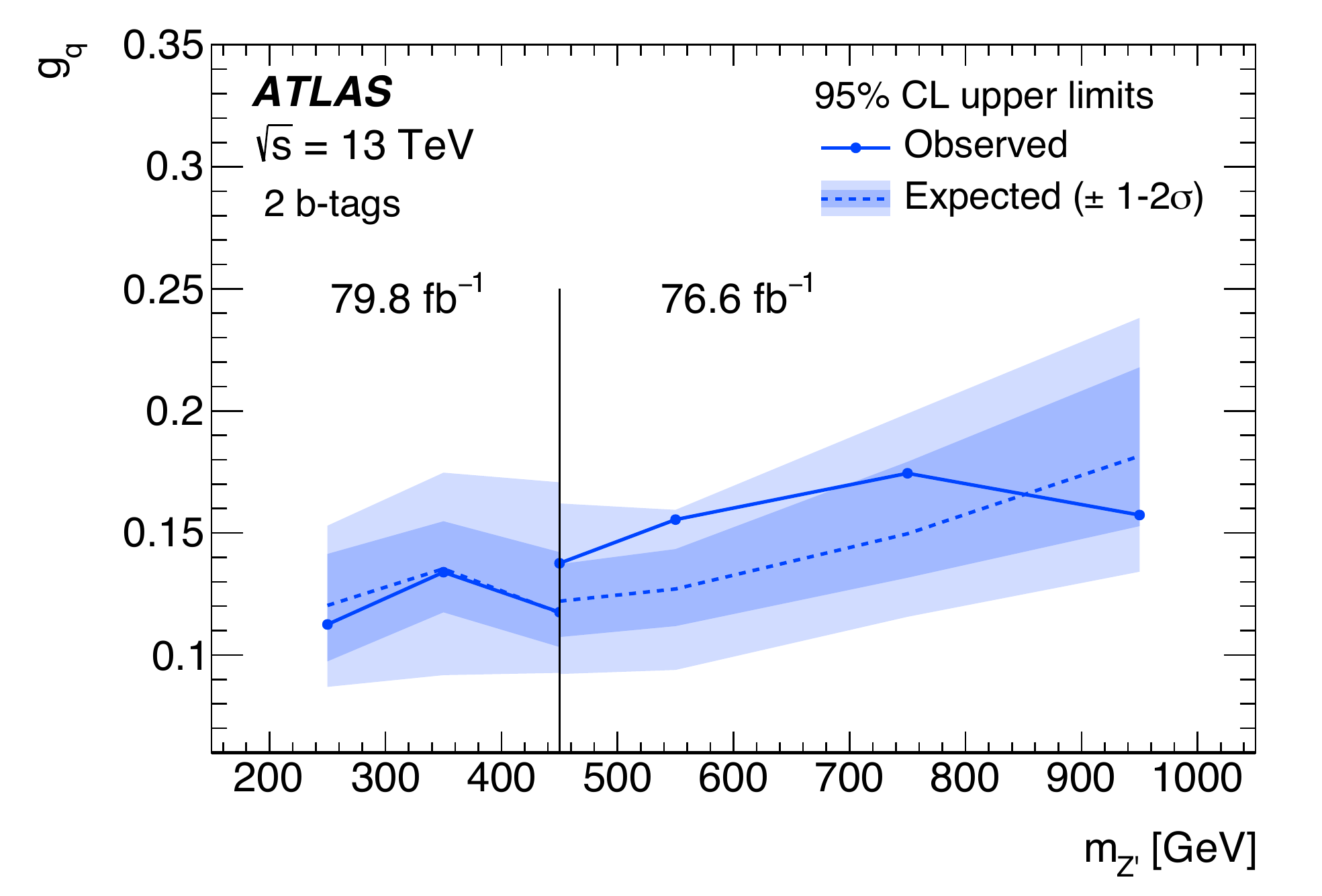}
\caption{\label{fig:limit_single_btag_0p2}}
 
\end{subfigure} \\
\caption[Upper limits on $\Zprime$ contributions]{
Excluded values of the coupling between a $\Zprime$ and quarks, at 95\% CL, as a function of $m_{\Zprime}$, from \subref{fig:limit_single_inc_0p2} the flavour-inclusive and \subref{fig:limit_single_btag_0p2} the \btagged categories.
Below $450\,\GeV$ the distribution of events selected by the single-photon trigger is used for hypothesis testing, while above $450\,\GeV$ the combined trigger is used.} \label{fig:limits_zprime}
\end{figure}
 
\begin{figure}
\centering
\begin{subfigure}[b]{0.49\textwidth}
\includegraphics[width=\textwidth]{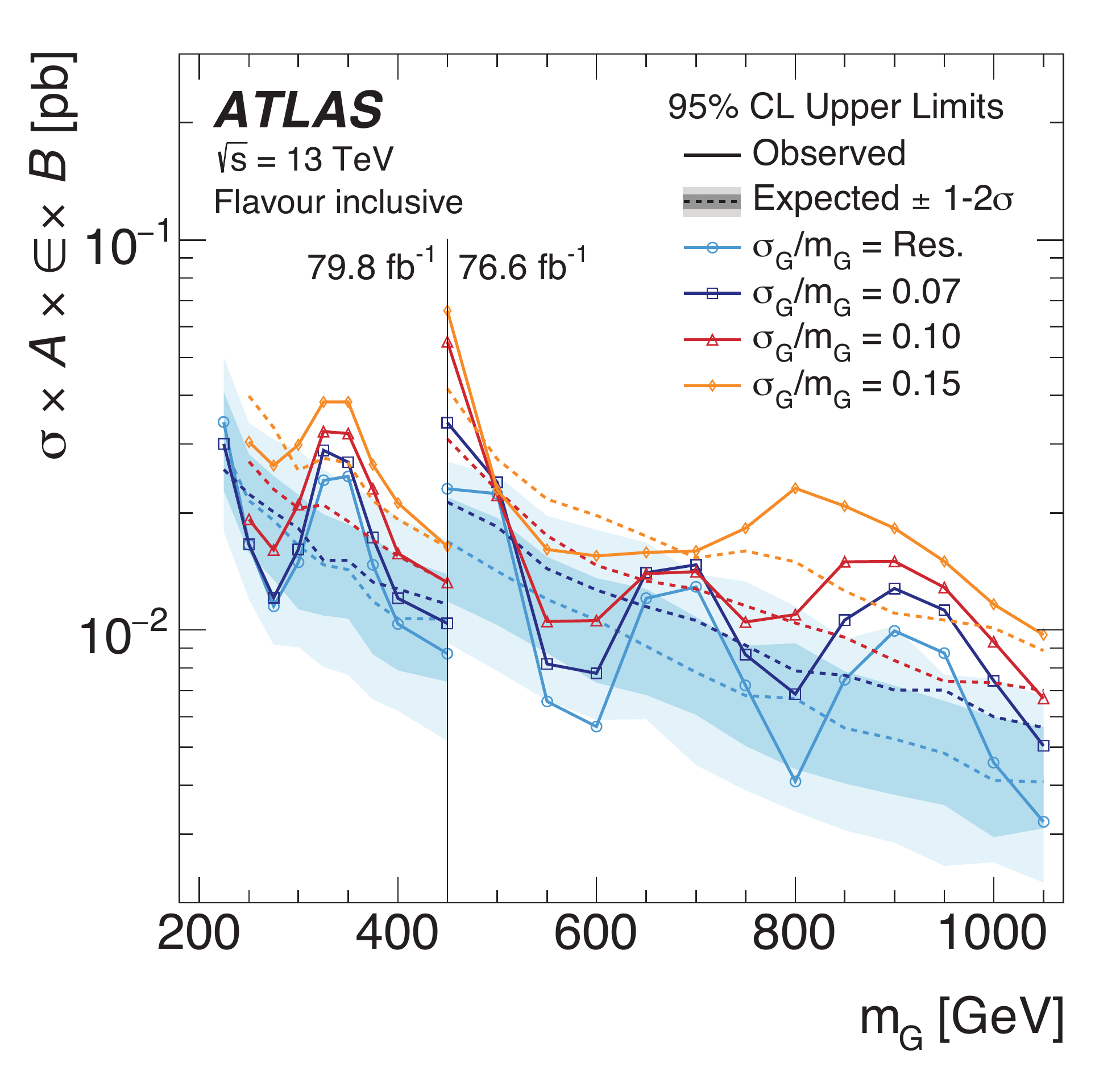}
\caption{\label{fig:limit_gaussian_inc}}
\end{subfigure}
\begin{subfigure}[b]{0.49\textwidth}
\includegraphics[width=\textwidth]{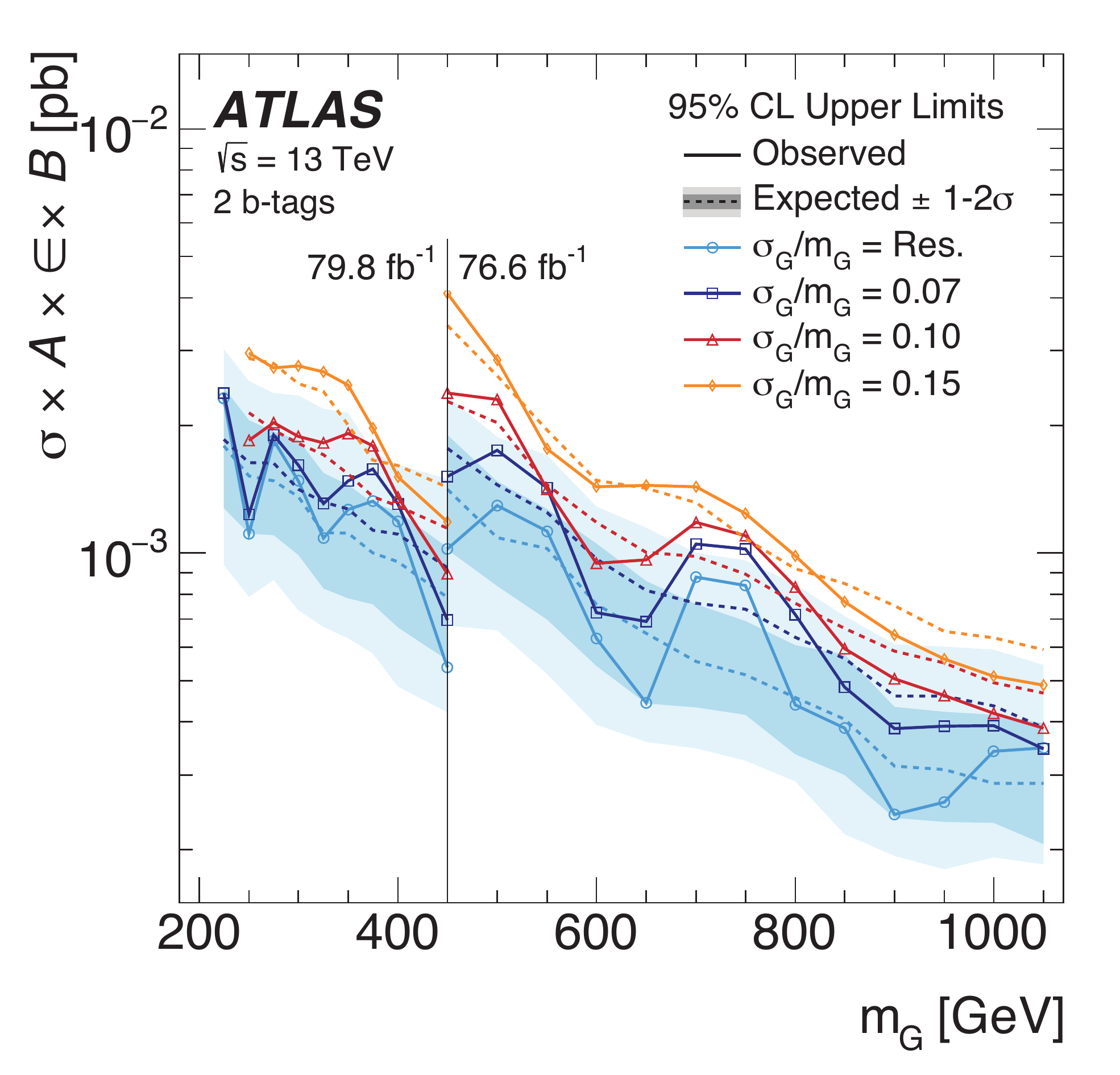}
\caption{\label{fig:limit_gaussian_btag}}
\end{subfigure}
\caption[]{Upper limits on Gaussian-shape contributions to the dijet mass distributions from \subref{fig:limit_gaussian_inc} the flavour-inclusive and \subref{fig:limit_gaussian_btag} the \btagged categories.
The curve denoted ``Res.'' represents the limit on intrinsically narrow contributions with Gaussian mass resolution ranging from 8\% to 3\% for the mass range considered.
Below $450\,\GeV$, the distribution of events selected by the single-photon trigger is used for hypothesis testing, while above $450\,\GeV$ the combined trigger category is used.
While the vertical axis is shared between the two selections, the signal acceptance is not the same below and above the line, and this results in different limits for the $450\,\GeV$ resonance mass point.
Thus the two sets of limit points correspond to two different interpretations of the product of cross-section, acceptance, efficiency, and branching ratio, $\sigma \times A \times \epsilon \times \mathcal{B}$.}
\label{fig:limits_gaussian}
\end{figure}

A more generic set of limits is shown in Fig.~\ref{fig:limits_gaussian}. These limits apply to the visible cross-section from a Gaussian-shape contribution to the $\mjj$ distribution, where the visible cross-section is defined as the product of the production cross-section, the detector acceptance, the reconstruction efficiency, and the branching ratio, $\sigma \times A \times \epsilon \times \mathcal{B}$.
The Gaussian-shape contributions have mass $m_\mathrm{G}$ and widths that span from the detector mass resolution, denoted ``Res.'' in the figure, ranging from 8\% to 3\% for the mass range considered, for an intrinsically narrow resonance, up to 15\% of the mean of the Gaussian mass distribution.

Both the choice of fit function and statistical fluctuations in the $\mjj$ distribution can contribute to uncertainties in the background model.
To account for the fit function choice, the largest difference between fits among the variants of Eq.~(\ref{eq:ua2}) and~ Eq.~(\ref{eq:5par}) that obtain a $p$-value above 0.05, is taken as a systematic uncertainty.
The uncertainty related to statistical fluctuations in the background model is computed via Poisson fluctuations around the values of the nominal background model.
The uncertainty of the prediction in each $\mjj$ bin is taken to be the standard deviation of the predictions from all random samples.
 
The reconstructed signal mass distributions are affected by additional uncertainties related to the simulation of detector effects.
The jet energy scale uncertainty is applied to the $\Zprime$ mass distributions using a four-principal-component method~\cite{PERF-2016-04,ATL-PHYS-PUB-2015-014,ATL-PHYS-PUB-2015-015}, leading to an average 2\% shift of the peak value for each mass distribution.
For the Gaussian-shape signal models, this average 2\% shift is taken as the uncertainty of the mean of each Gaussian distribution.
In the case of the \btagged\ categories, uncertainties of the \btagging\ efficiency are the dominant uncertainties in each mass distribution.
To account for these uncertainties, the contribution of each simulated event to a given mass distribution is reweighted by 5\%--15\% for each jet, depending on its $\pt$~\cite{PERF-2016-05}.
 
The remaining uncertainties are modelled by scaling each simulated distribution by 3\% to account for jet energy resolution in all categories~\cite{PERF-2016-04}, 2\% for photon identification uncertainties in the single-photon-trigger categories and 1.4\% in the combined-trigger categories~\cite{PERF-2017-02}, 3\% to account for efficiencies of the combined trigger, and 1\% for PDF-related uncertainties (only applied to the mass distributions of $\Zprime$ signals).
 
All these uncertainties are included in the reported limits; further uncertainties of the theoretical cross-section for the \Zprime model are not considered.
 
The uncertainty of the combined 2015--2017 integrated luminosity is derived following a methodology similar to that detailed in Ref.~\cite{DAPR-2013-01} and using the LUCID-2 detector for the baseline luminosity measurements in 2017~\cite{LUCID2}.
The estimates for the individual datasets are combined and applied as a single scaling parameter with a value of 2\% for the single-photon-trigger categories and 2.3\% for the combined-trigger categories.
 
\section{Conclusion}
 
Dijet resonances with a width up to 15\% of the mass, produced in association with a photon, were searched for in up to $79.8\,\ifb$ of LHC $pp$ collisions recorded by the ATLAS experiment at $\sqrt{s}=13\,\TeV$. The observed $\mjj$ distribution in the mass range $169\,\GeV < \mjj < 1100\,\GeV$ can be described by a fit with smooth functions without contributions from such resonances.

In the absence of a statistically significant excess, limits are set on two models: $\Zprime$ axial-vector dark-matter mediators and Gaussian-shape signal contributions.
All mediator masses within the analysis range are excluded for a coupling value of $\gq = 0.25$ and above, with the exclusion limit near a coupling of $\gq=0.15$ for most of the mass range.
The \btagged categories yield $\Zprime$ limits comparable to the flavour-inclusive categories, assuming that the $\Zprime$ decays equally into all quark flavours, and provide model-independent limits that can be reinterpreted in terms of resonances decaying preferentially into $b$-quarks.
For narrow Gaussian-shape structures with a width-to-mass ratio of 7\%, the flavour-inclusive categories exclude visible cross-sections above 12~fb for a mass of 400~\GeV and above 5.1~fb for a mass of 1050~\GeV.
When wider signals with a width-to-mass ratio of 15\% are considered, the exclusion limits are weaker at the lower mass values, with visible cross-sections above 21~fb excluded for a mass of $400\,\GeV$ and those above 9.7~fb excluded for a mass of $1050\,\GeV$.
 
These results significantly extend the constraints by ATLAS and other experiments at lower centre-of-mass energies on hadronically decaying resonances with masses as low as 225 GeV and up to 1100 GeV.
 
\section*{Acknowledgements}
 
 
We thank CERN for the very successful operation of the LHC, as well as the
support staff from our institutions without whom ATLAS could not be
operated efficiently.
 
We acknowledge the support of ANPCyT, Argentina; YerPhI, Armenia; ARC, Australia; BMWFW and FWF, Austria; ANAS, Azerbaijan; SSTC, Belarus; CNPq and FAPESP, Brazil; NSERC, NRC and CFI, Canada; CERN; CONICYT, Chile; CAS, MOST and NSFC, China; COLCIENCIAS, Colombia; MSMT CR, MPO CR and VSC CR, Czech Republic; DNRF and DNSRC, Denmark; IN2P3-CNRS, CEA-DRF/IRFU, France; SRNSFG, Georgia; BMBF, HGF, and MPG, Germany; GSRT, Greece; RGC, Hong Kong SAR, China; ISF and Benoziyo Center, Israel; INFN, Italy; MEXT and JSPS, Japan; CNRST, Morocco; NWO, Netherlands; RCN, Norway; MNiSW and NCN, Poland; FCT, Portugal; MNE/IFA, Romania; MES of Russia and NRC KI, Russian Federation; JINR; MESTD, Serbia; MSSR, Slovakia; ARRS and MIZ\v{S}, Slovenia; DST/NRF, South Africa; MINECO, Spain; SRC and Wallenberg Foundation, Sweden; SERI, SNSF and Cantons of Bern and Geneva, Switzerland; MOST, Taiwan; TAEK, Turkey; STFC, United Kingdom; DOE and NSF, United States of America. In addition, individual groups and members have received support from BCKDF, the Canada Council, CANARIE, CRC, Compute Canada, FQRNT, and the Ontario Innovation Trust, Canada; EPLANET, ERC, ERDF, FP7, Horizon 2020 and Marie Sk{\l}odowska-Curie Actions, European Union; Investissements d'Avenir Labex and Idex, ANR, R{\'e}gion Auvergne and Fondation Partager le Savoir, France; DFG and AvH Foundation, Germany; Herakleitos, Thales and Aristeia programmes co-financed by EU-ESF and the Greek NSRF; BSF, GIF and Minerva, Israel; BRF, Norway; CERCA Programme Generalitat de Catalunya, Generalitat Valenciana, Spain; the Royal Society and Leverhulme Trust, United Kingdom.
 
The crucial computing support from all WLCG partners is acknowledged gratefully, in particular from CERN, the ATLAS Tier-1 facilities at TRIUMF (Canada), NDGF (Denmark, Norway, Sweden), CC-IN2P3 (France), KIT/GridKA (Germany), INFN-CNAF (Italy), NL-T1 (Netherlands), PIC (Spain), ASGC (Taiwan), RAL (UK) and BNL (USA), the Tier-2 facilities worldwide and large non-WLCG resource providers. Major contributors of computing resources are listed in Ref.~\cite{ATL-GEN-PUB-2016-002}.
 

\clearpage
\printbibliography
\clearpage 
 
\begin{flushleft}
{\Large The ATLAS Collaboration}

\bigskip

M.~Aaboud$^\textrm{\scriptsize 34d}$,    
G.~Aad$^\textrm{\scriptsize 100}$,    
B.~Abbott$^\textrm{\scriptsize 126}$,    
D.C.~Abbott$^\textrm{\scriptsize 101}$,    
O.~Abdinov$^\textrm{\scriptsize 13,*}$,    
A.~Abed~Abud$^\textrm{\scriptsize 69a,69b}$,    
D.K.~Abhayasinghe$^\textrm{\scriptsize 92}$,    
S.H.~Abidi$^\textrm{\scriptsize 165}$,    
O.S.~AbouZeid$^\textrm{\scriptsize 39}$,    
N.L.~Abraham$^\textrm{\scriptsize 154}$
% ...
\bigskip
\\

% ...
$^{13}$Institute of Physics, Azerbaijan Academy of Sciences, Baku; Azerbaijan.\\
% ...
$^{34}$$^{(a)}$Facult\'e des Sciences Ain Chock, R\'eseau Universitaire de Physique des Hautes Energies - Universit\'e Hassan II, Casablanca;$^{(b)}$Centre National de l'Energie des Sciences Techniques Nucleaires (CNESTEN), Rabat;$^{(c)}$Facult\'e des Sciences Semlalia, Universit\'e Cadi Ayyad, LPHEA-Marrakech;$^{(d)}$Facult\'e des Sciences, Universit\'e Mohamed Premier and LPTPM, Oujda;$^{(e)}$Facult\'e des sciences, Universit\'e Mohammed V, Rabat; Morocco.\\
% ...
$^{39}$Niels Bohr Institute, University of Copenhagen, Copenhagen; Denmark.\\
% ...
$^{69}$$^{(a)}$INFN Sezione di Pavia;$^{(b)}$Dipartimento di Fisica, Universit\`a di Pavia, Pavia; Italy.\\
% ...
$^{92}$Department of Physics, Royal Holloway University of London, Egham; United Kingdom.\\
% ...
$^{100}$CPPM, Aix-Marseille Universit\'e, CNRS/IN2P3, Marseille; France.\\
$^{101}$Department of Physics, University of Massachusetts, Amherst MA; United States of America.\\
% ...
$^{126}$Homer L. Dodge Department of Physics and Astronomy, University of Oklahoma, Norman OK; United States of America.\\
% ...
$^{154}$Department of Physics and Astronomy, University of Sussex, Brighton; United Kingdom.\\
% ...
$^{165}$Department of Physics, University of Toronto, Toronto ON; Canada.\\
% ...
$^{a}$ Also at Borough of Manhattan Community College, City University of New York, NY; United States of America.\\
$^{b}$ Also at California State University, East Bay; United States of America.\\
$^{c}$ Also at Centre for High Performance Computing, CSIR Campus, Rosebank, Cape Town; South Africa.\\
$^{d}$ Also at CERN, Geneva; Switzerland.\\
$^{e}$ Also at CPPM, Aix-Marseille Universit\'e, CNRS/IN2P3, Marseille; France.\\
% ...
$^{*}$ Deceased

\end{flushleft}


\end{document}